# Structural, Optical and Electronic properties of Magnetron Sputtered Si and Ge thin films

**A. Dutta[+, *], S. Tripathi[†], R. Brajpuriya[†], A. Sharma[†] and T. Shripathi[†]**

[†]UGC-DAE Consortium for scientific research, University campus, Khandwa road, Indore-452017, India

[+]Dept. of Physics, Indian Institute of Technology, Dhanbad

Email: [*]aniruddha@knights.ucf.edu

*Abstract*

*We report the results of GIXRR, UV-VIS NIR and XPS measurements on Si and Ge thin films of various thicknesses. While GIXRR measurements show no presence of oxide on the top of these films, XPS measurements show small amount of oxides. Also, a sharp increase in surface roughness is seen with thickness in agreement with the columnar growth of films. In contrast, in case of Si films, oxide is present even inside the layers but in a very small amount and is not detected in GIXRR/absorption measurements.*

## INTRODUCTION

A range of investigations on semiconductor materials including fundamental and applied properties have been initiated in last few years [1]. Typical structures consisting can be used to engineer their properties, which may be the basis of next generation tuneable band gap structures /devices. Si and Ge both have similar properties and applications and are currently the materials most used in semiconductor industry, particularly in logic and memory applications. This has motivated intense research aimed at the materials with well-controlled composition and nanoscale structure. Ultimate control over the properties of these materials has therefore become a major theme in material science research [2]. Therefore, present work is focused on the study of magnetron sputtered Si and Ge thin films using various characterization techniques.

## EXPERIMENTAL

Si and Ge films Ge-I (Si-I), Ge-II (Si-II) and Ge-III (Si-III) were deposited by Magnetron Sputtering on float glass at high vacuum. GIXRR measurements were done on Bruker D8 Discover diffractometer with Cu source ($\lambda$=1.54Å). XPS measurements were done on VSW XPS system (Al K$\alpha$). All UV-VIS NIR spectroscopy measurements were performed using Perkin-Elmer Lambda-950 dual beam spectrophotometer.

## RESULTS AND DISSCUSION

**GIXRR measurements:**

Figure 1(i) shows the GIXRR patterns for Ge-I, II, & III. The reflectivity remains maximum upto the critical angle of ~0.30 for all films, matching with bulk Ge. Film thicknesses calculated using Parratt formulism [3] are given in table-1. It is clear that electron densities are close to bulk value, showing that Ge film grows in a dense manner. As we move from Ge-I to Ge-II, the number of oscillations increases and they come closer. Also, the roughness drastically increases in Ge-II because after a few atomic layers, Ge grows in a columnar pattern and may cause observed increase. For Ge-III, denser oscillations are seen.

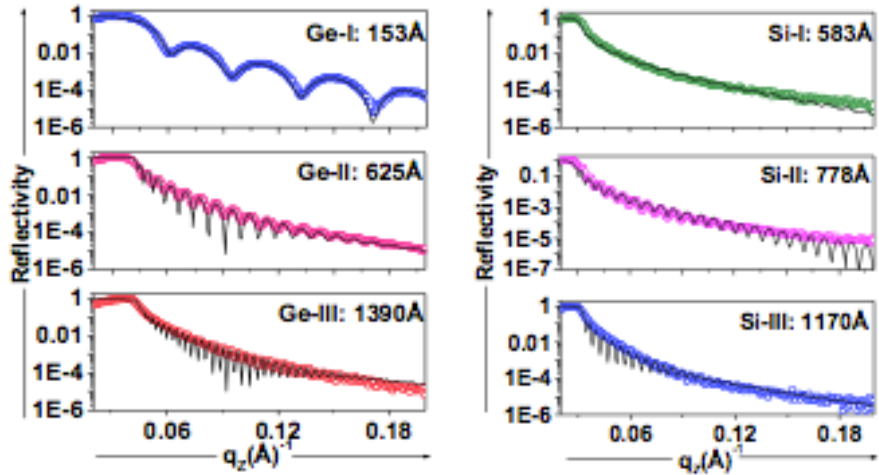

**Figure 1:** GIXRR patterns of (i) Ge and (ii) Si thin films

Similarly, for Si-I, II, III, critical angle is close to bulk value of ~0.22º [figure 1(ii)] and electron density is slightly less than corresponding bulk Si due to atomic rearrangement taking place at the film/substrate interface. It leads to the possibility that many vacancies and point defects incorporated inside the layers.

| Sample | | Thickness (Å) | Roughness (Å) | Elect. density (1/Å$^2$) |
|---|---|---|---|---|
| **Standard Ge bulk** | | ---- | ---- | 3.843E-05 |
| **Ge Thin film** | I | 152.5 | 7.807 | 3.82E-05 |
| | II | 625 | 16.356 | 3.632E-05 |
| | III | 1390 | 16.052 | 3.839E-05 |
| **Standard Si bulk** | | ---- | ---- | 2.012E-05 |
| **Si Thin film** | I | 583 | 7.031 | 1.907E-05 |
| | II | 778 | 9.219 | 1.556E-05 |
| | III | 1169 | 22.379 | 1.167E-05 |

**Table 1:** GIXRR parameters for Ge and Si thin films

Also, roughness is increasing resulting in increased number of defects. This results in even lower electron density. Similar to Ge, this fact is attributed to columnar growth of Si films.

**XPS measurements:**

Figure 2(i) shows the XPS survey scan for Ge-II. It can be seen that apart from carbon (C) and oxygen (O), no other impurity elements are present confirming the purity of the sample. After sputtering with Ar$^+$ ions for 20 min, C and O are almost eliminated and Ge peaks increase in intensity with clear appearance of Ge Auger peaks. On further sputtering for 40 min only clean Ge film is seen with almost no traces of C and O [see inset].

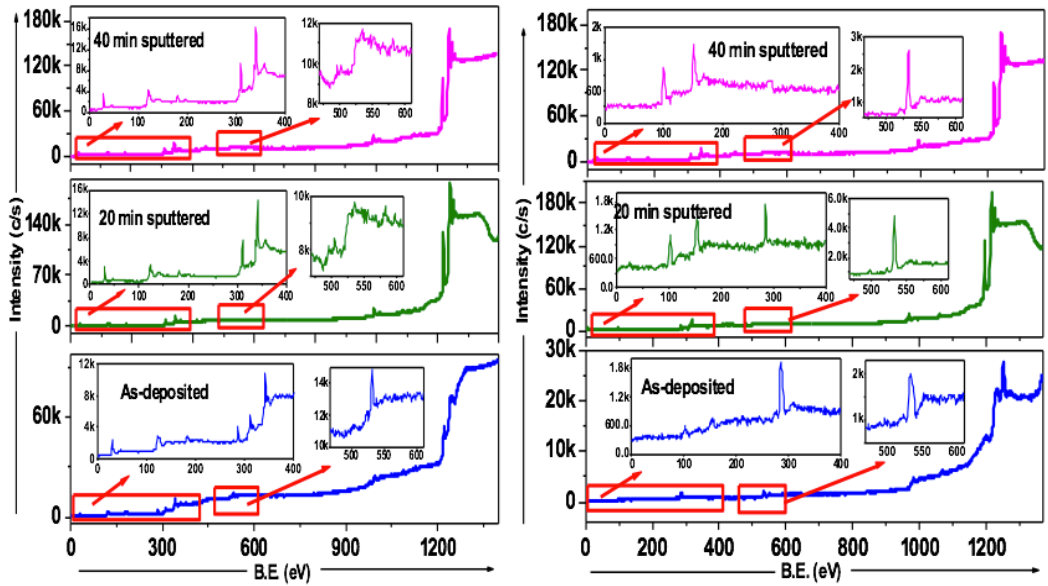

**Figure 2**: XPS survey scans of (i) Si and (ii) Ge films

The survey scan of Si-II film is shown in Figure 2(ii). After 20 min sputtering, C reduces while O enhances. However, after 40 min sputtering, C is almost removed with significant presence of O peak. On comparing these results with Ge film, we can say that Si has more affinity for O than Ge and it may happen that during deposition along with Si atoms, the O atoms are also incorporated.

**UV-Vis Measurements:**

Figure 3(i) depicts %Transmission (T) for Ge films. The spectra show only smooth absorption bands indicating amorphous grains. The absorption edges shift from Ge-I to Ge-III probably due to slight differences in atomic densities. With increasing thickness, Ge-III shows comparatively sharp edge with long tail towards higher hν, whole in low hν region, the films are highly transparent with %T decreasing with thickness except for the highest thickness film.
Similarly, for Si films, %T-hν curves follow the shift towards lower hν with increasing thickness [figure 3(ii)]. Also, the gradual fall changes into a comparatively sharp edge. In the low hν region, all films are more transparent than even the lowest thickness Ge films may be due to lower density of Si as compared to Ge. The gradual onset of absorption suggests that an energy broadening of electronic states occurs in amorphous Si films. For indirect band gap materials, the absorption coefficient ($\alpha$) depends on hν via the relation: $\alpha = a (h\nu - E_g)^2 / h\nu$, where a is constant and $E_g$ is bandgap.

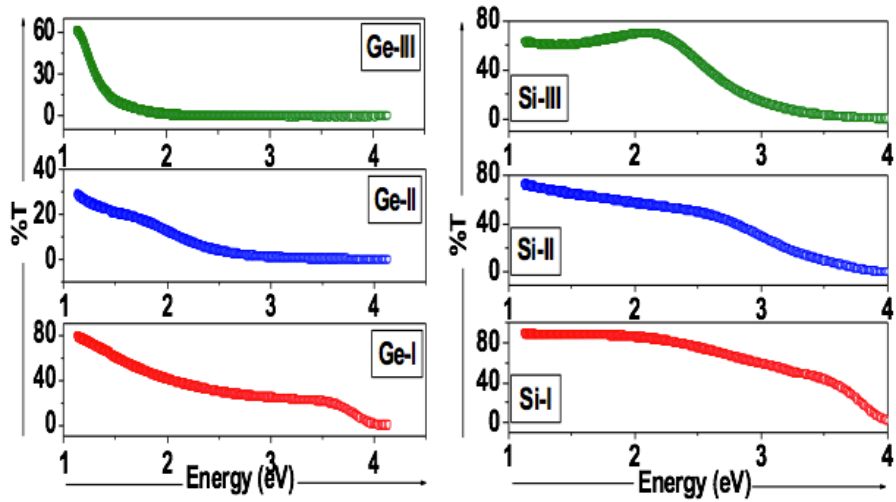

**Figure 3:** (i, ii) %T vs hν plots at λ=300-1100nm of Ge and Si thin films respectively

Therefore, a clearer picture can be obtained by plotting $\sqrt{(\alpha h\nu)}$ vs. hν [figure 4(i), (ii)].

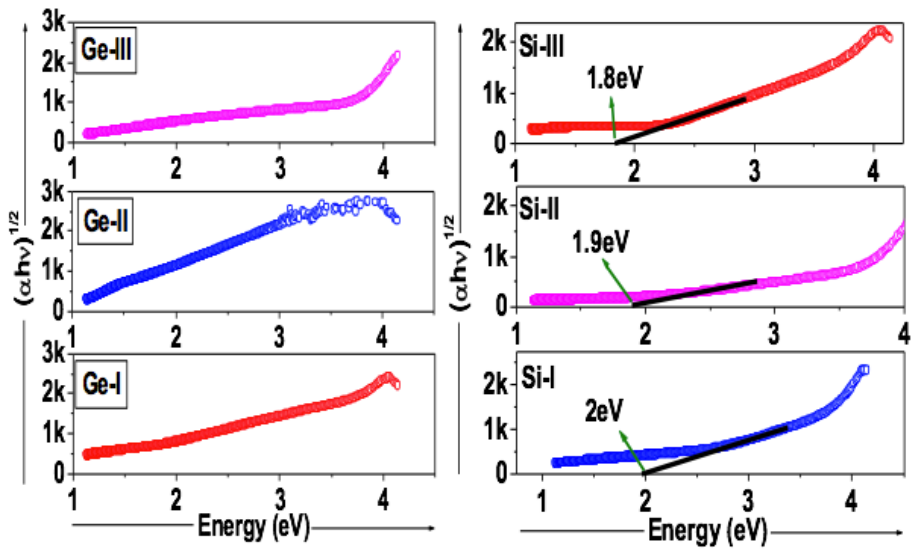

**Figure 4:** $(\alpha h\nu)^{1/2}$ vs hν plots of (i) Ge and (ii) Si thin films

In case of all Ge films, no sharp transition is observed and only a long Urbach Tail is seen. It is not possible to calculate the band gap of Ge films due to their amorphous nature. However, the band gaps of Si films match with reported values [4]. With increased film thickness as we approach bulk side, there is a slight decreasing trend in band gap. From XPS results some amount of $SiO_2$ was seen inside Si films, therefore the effect should also be observed in absorption measurements. Pure $SiO_2$ behaves like an insulator and its absorption edge occurs at ~7eV. But in the present case the edge is at much lower hν (1-4eV) confirming our earlier observation that $SiO_2$ is present in a negligible amount and does not affect the overall absorption behaviour further suggesting that XPS is much more surface sensitive than GIXRR/absorption measurements.

# CONCLUSION

The structural, optical and electronic properties of Ge and Si thin films have been investigated. GIXRR measurements show no presence of oxide with sharp increase in surface roughness with thickness in agreement with the columnar growth of films. In contrast, XPS measurements show small amount of oxides on top surface. In case of Si films, very small amount of oxide is present even inside the layers and is not detected in GIXRR/absorption measurements. The films are amorphous in nature.

# ACKNOWLEGEMENTS

We thank Prof A. Gupta (GIXRR), Dr. D. M. Phase (Deposition), and Mr. U. Deshpande (PES and UV-VIS NIR), UDCSR, Indore for their help. The authors (ST, RB, AS) are thankful to CSIR for providing financial support.